\documentclass{ws-ijmpcs}

\begin{document}

\markboth{Martin Lemoine}
{Relativistic shock acceleration and some consequences}

%
\catchline{}{}{}{}{}
%

\title{Relativistic shock acceleration and some consequences}

\author{Martin Lemoine}

\address{Institut d'Astrophysique de Paris, CNRS -- UPMC, 98 bis
  boulevard Arago\\
F-75014, Paris, France\\
lemoine@iap.fr}

\author{Guy Pelletier}

\address{Institut de Plan\'etologie et Astrophysique de Grenoble, CNRS
  -- UJF, 214 rue de la Piscine\\
F-38041 Grenoble Cedex, France\\
guy.pelletier@obs.ujf-grenoble.fr}

\maketitle

\begin{history}
\received{Day Month Year}
\revised{Day Month Year}
\end{history}

\begin{abstract}
This paper summarizes recent progresses in our theoretical
understanding of particle acceleration at relativistic shock waves and
it discusses two salient consequences: (1) the maximal energy of
accelerated particles; (2) the impact of the shock-generated
micro-turbulence on the multi-wavelength light curves of gamma-ray
burst afterglows.

\keywords{shock waves; particle acceleration; gamma-ray bursts}
\end{abstract}

\ccode{PACS numbers: 11.25.Hf, 123.1K}

\section{Introduction}	
The physics of particle acceleration at relativistic shock waves plays
a central role in the modelling of various powerful astrophysical
sources, e.g. micro-quasars, pulsar wind nebulae, gamma-ray bursts and
active galactic nuclei. In the absence of well motivated
prescriptions, it has been customary to adopt phenomenological
scalings for the acceleration timescale, such as a Bohm-like $t_{\rm
  acc}\,\simeq\,{\cal A}\,t_{\rm g}$ (with $t_{\rm g}$ the gyro-time,
${\cal A}$ a fudge factor) in order to compute quantities of interest,
in particular the maximal energy at acceleration.

In the past decades, however, our theoretical understanding of shock
physics and shock acceleration has made substantial progress, to the
point where one can start to make definite predictions on the inner
mechanics of the acceleration process. This contribution to HEPRO-IV
summarizes some of these achievements and it discusses some salient
consequences for high energy astrophysics.

\section{Shock acceleration and micro-physics}
A central point in the present discussion is the realization that the
development of Fermi-like acceleration at ultra-relativistic
collisionless shocks -- $\gamma_{\rm sh}\beta_{\rm sh}\,\gg\,1$ with
$\beta_{\rm sh}\,=\,(1-1/\gamma_{\rm sh}^2)^{1/2}$ the shock velocity
in units of $c$ -- is intimately connected to the generation of
micro-turbulence, on a scale $\lambda_{\delta B}\,\ll\,r_{\rm
  g}\,\equiv\,c\,t_{\rm g}$. This has come from different point of
views:
\begin{itemize}
\item Motivated by the modelling of gamma-ray burst afterglows, which
  seemingly pointed to the existence of a magnetic field close to
  equipartition in the shocked region, with
  $\epsilon_B\,\sim\,10^{-2}$ -- $\epsilon_B\,=\,\delta
  B^2/\left(16\pi\gamma_{\rm sh}^2n_{\rm u}m_pc^2\right)$, with
  $n_{\rm u}$ the proper density of the unshocked plasma, denotes the
  equipartition fraction of the turbulent magnetic field downstream of
  the shock -- well above the shock compressed interstellar magnetic
  field value, Refs.~\cite{GW99,ML99} have proposed that the magnetic
  field was self-generated in the shock precursor through
  micro-instabilities, in particular the Weibel/filamentation
  mode. Such instabilities naturally produce turbulence on microscopic
  skin depth scales $c/\omega_{\rm pi}\,\sim\,10^7\,n_{\rm u,0}\,$cm
  downstream of the shock. Furthermore, for shock heated particles of
  energy $\epsilon\,\sim\,\gamma_{\rm sh}m_pc^2$, it is easy to see
  that $r_{\rm g}\,\sim\,\epsilon_B^{-1/2}\lambda_{\delta
    B}\,\gtrsim\,\lambda_{\delta B}$. From the point of view of the
  accelerated particle population, the turbulence thus lies on small
  scales.

\item Up to ten years ago, most studies of particle acceleration at
  relativistic shock waves relied on a test-particle picture, in which
  one treats the shock as a discontinuity and one ignores the
  back-reaction of the accelerated particles on the shock
  environment. However, detailed analyses of the particle kinematics,
  through theoretical arguments~\cite{LPR06} and Monte Carlo
  simulations~\cite{NOP06}, have revealed that the Fermi process can
  develop only if intense small scale turbulence has been excited on
  scales smaller than $r_{\rm g}$. Otherwise, the particle would be
  advected with the flow at velocity $c/3$, away from the shock,
  before it has time to scatter back across the magnetic field; the
  latter is indeed essentially perpendicular downstream of the shock,
  due to Lorentz transform and compression effects (i.e.  a so-called
  superluminal configuration~\cite{BK90}).

  As noted in Ref.~\cite{LPR06}, this argument offers an interesting
  connection with phenomenology and observations: if the turbulence in
  gamma-ray burst blast waves is indeed of Weibel/filamentation
  origin, one has $\lambda_{\delta B}\,<\,r_{\rm g}$ hence this
  turbulence should provide the requisite conditions for relativistic
  Fermi acceleration. This point of view has been confirmed by recent
  particle-in-cell (PIC) simulations, see below.

  Further work has pinpointed the requisite characteristics to
  accelerate particles via the ultra-relavistic Fermi
  process~\cite{PLM09,LP10}: $\lambda_{\delta B}\,\lesssim\,r_{\rm
    g}\,\lesssim\,\lambda_{\delta B}\,\delta B/B$ with $B$ the
  background field seen in the downstream rest frame
  ($B=\sqrt{8}\gamma_{\rm sh}B_{\rm u}$ for a perpendicular shock,
  $B_{\rm u}$ the upstream-frame field). Note that the r.h.s. imposes
  an upper bound on the energy of the particle: as energy grows, the
  scattering frequency in the micro-turbulent field $\nu_{\rm
    s}\,\sim\,c\lambda_{\delta B}/r_{\rm g}^2$ becomes less
  competitive compared to the gyration frequency $c/r_{\rm g}$,
  therefore scattering eventually becomes ineffective and the particle
  is dragged away from the shock in the perpendicular magnetic
  field~\cite{PPL11}.  One can rewrite the above in terms of the
  equipartition fraction parameter $\epsilon_B$ and the magnetization
  parameter $\sigma\,\equiv\, B_{\rm u}^2/\left(4\pi n_{\rm u}m_p
  c^2\right)$ as follows~\cite{LP10}:
\begin{equation}
\sigma \,\lesssim\,\epsilon_B^2\ ,\label{eq:acc1}
\end{equation}
where $\epsilon_B$ should be understood in this equation as the
average of this value downstream of the shock. Hence, from a purely
theoretical analysis of the accelerated particle kinematics, one can
predict that acceleration will take place in weakly magnetized shock
waves $\sigma\,\ll\,1$. The same arguments teach us that the
scattering must take place at a rate $\nu_{\rm
  s}\,\propto\,\epsilon^{-2}$. As we will see in the following, these
scalings have also been confirmed by particle-in-cell (PIC)
simulations.

\item Finally, the advent of HPC PIC simulations have shown that the
  micro-turbulence is itself an integral part of the collisionless
  shock~\cite{S05,K07,S08a,Nea09}: micro-instabilities akin to the
  Weibel/filamentation mode develop in the shock precursor and
  build-up the micro-turbulence on skin depth scales; this turbulence
  then grows to a sub-equipartition fraction $\epsilon_B\,\sim\,0.1$
  close the shock front, whereby it builds an electromagnetic barrier
  which isotropizes the incoming background plasma (as seen in the
  shock front rest-frame), i.e. it initiates the shock transition.
\end{itemize}
The three above point of views meet and complement each other,
underlying the intimate non-linear relationship between shock
structure, micro-instabilities and particle acceleration.

The simulations of Ref.~\cite{S08b} have made another step forward, in
demonstrating that in the unmagnetized limit $\sigma\,\rightarrow\,0$,
the self-generated micro-turbulence provides the source of scattering
that leads to the development of the Fermi process. The number of PIC
simulations of shock acceleration has then kept growing, in order to
probe in which conditions of magnetization and shock velocity
acceleration takes place and at what rate; such progresses are related
in the contribution of L. Sironi in this volume. These simulations
confirm the predictions of the theoretical analyses described above:
in particular, Fermi acceleration is seen to take place at low
magnetization $\sigma\,\lesssim\,10^{-5}$, as expected from
Eq.~\ref{eq:acc1}, with a rate of order $\nu_{\rm s}$, also as
expected.

Naturally, the study of micro-instabilities upstream of shock fronts
has received ample attention, in particular the dominant
Weibel/filamentation
mode~\cite{ML99,WA04,LE06,AW07,AWN07,BGB10,LP10,LP11,Rea11,Sea12} at
low magnetization levels. The comparison of the growth rate of this
instability, $\Im\omega\,\sim\,\xi_{\rm cr}^{1/2}\omega_{\rm pi}$
(with $\xi_{\rm cr}\,\sim\,0.1$ the fraction of incoming energy
transferred to the accelerated particle population) with the rate at
which the background plasma crosses the shock precursor,
$t_{\times}^{-1}\,\sim\, \sigma^{1/2}\gamma_{\rm sh}\omega_{\rm pi}$,
limits the parameter space in which this instability actually has time
to grow and build up the micro-turbulence~\cite{LP10}:
\begin{equation}
\sigma\,\lesssim\,\gamma_{\rm sh}^{-2}\xi_{\rm cr}\ .\label{eq:wf}
\end{equation}
Whenever this bound is not satisfied, the Weibel/filamentation
instability cannot structure the shock by itself. As discussed by
G.~Pelletier in this volume, the relevant instability over all
remaining parameter space, is a current-driven instability which can
develop for any value of the shock Lorentz factor, provided
$\sigma\,\lesssim\,10^{-2}$. This instability also produces
micro-turbulence on a skin depth scale, therefore the results that
follow apply to this case as well.  Above $10^{-2}$ in magnetization,
the shock can be mediated by the direct compression of the magnetic
field and by the development of the synchrotron-maser
instability~\cite{AA88}.

In spite of these successes, one must keep in mind the rather large
number of open questions in this field, as well as the gap which
remains between the timescales probed by the PIC simulations and the
astrophysical timescales. So far, the longest PIC
simulation~\cite{Kea09} has run over $\sim\,10^4\,\omega_{\rm p}^{-1}$
for a pair shock, which represents a fraction of a percent of the
dynamical timescale of a gamma-ray burst in the early afterglow
phase. Theoretical extrapolation is thus needed to connect these
simulations to actual sources. Finally, as emphasized in
Ref.~\cite{Kea09}, PIC simulations have not yet converged to a
stationary state and the acceleration of particles to progressively
higher energies appears to feed back on the shock structure as time
goes on. With these caveats in mind, we now discuss two macrophysical
consequences of this microphysics of relativistic shock acceleration.

\section{Acceleration to very high energies}
Formally, the acceleration timescale in the shock frame reads: $t_{\rm
  acc}\,\simeq\,\left(t_{\rm d\vert sh} + t_{\rm u\vert sh}\right)/2$,
with $t_{\rm d\vert sh}$ (resp. $t_{\rm u\vert sh}$) the downstream
(resp. upstream) residence time expressed in the shock frame; the
factor $2$ corresponds to the typical energy gain per
cycle~\cite{Aea01,LP03}. These residence times depend inversely on the
scattering frequency, which in turn is controlled by the scale, the
velocity and geometry of the micro-turbulence, as discussed in detail
in ~\cite{PPL11,PPL13}. In particular, scattering in the shock
precursor may be hampered by the anisotropic nature of the turbulence,
since the near independence of turbulent modes along a given direction
implies the conservation of the conjugate momentum~\cite{Jea98}. In
such cases, the scattering may be controlled by the background
magnetic field upstream of the shock, but by the micro-turbulence
downstream of the shock, which appears essentially isotropic and
static there~\cite{Cea08,Kea09}.

\subsection{Electrons}
To simplify the discussion, one may assume that scattering takes place
in the micro-turbulence at a rate $\nu_{\rm
  s}\,\simeq\,c\lambda_{\delta B}/r_{\rm g}^2$ on both sides of the
shock. Then, balancing $t_{\rm acc}\,\sim\,\nu_{\rm s}^{-1}$ with the
timescale for synchrotron losses allows to compute the maximal Lorentz
factor of shock-accelerated electrons~\cite{KR10}:
\begin{equation}
\gamma_{e,\rm max}\,\simeq\,\left(r_e^3 n_{\rm
  u}m_e/m_p\right)^{-1/6}\,\simeq\, 7\times 10^6\,n_{\rm u,0}^{-1/6}\ ,
\end{equation}
in terms of the classical electron radius $r_e$, assuming here
$\lambda_{\delta B}\,\simeq\,c/\omega_{\rm p}$.  This gives rise to
synchrotron photons of maximal energy
\begin{equation}
\epsilon_{\gamma,\rm max}\,\simeq\, 3\,{\rm
  GeV}\,\,\epsilon_{B,-2}^{1/2}\gamma_{\rm sh,2.5}^2n_{\rm
  u,0}^{1/2}\ .
\end{equation}
In order to draw comparison with actual observations of gamma-ray
burst afterglows, one needs to take into account the temporal
dependence of the shock Lorentz factor through the deceleration phase;
the above maximal energy then corresponds to an observer timescale
$t_{\rm obs}\,\sim\,100\,$s for an isotropic shock energy
$10^{54}\,$ergs; beyond this timescale, it decays as $t_{\rm
  obs,2}^{-\alpha_\epsilon}$, with
$\alpha_\epsilon\,\simeq\,2/3\rightarrow 3/4$ depending on the
external density profile, see~\cite{L13,SSA13,WLL13}. 

The {\it Fermi-LAT} instrument has detected extended emission
$>100\,$MeV well beyond the prompt duration phase in a subset of
gamma-ray bursts; at high energies, the spectral power appears to
follow the scaling $F_\nu\,\propto \nu^{-\alpha}t_{\rm obs}^{-\beta}$
with $\alpha\,\sim\,1$ and $\beta\,\sim\,1$, as expected for the
synchrotron contribution of an electron spectrum ${\rm d}n_e/{\rm
  d}\gamma_e\,\propto \gamma_e^{-s}$ with $s\,\simeq\,2.2$. Very few
photons have been collected with source rest-frame energies
$>10\,$GeV, therefore it is not yet possible to constrain the maximal
energy at acceleration from observations. The above discussion rather
suggests that those photons seen above $10\,$GeV are most likely of
inverse Compton origin~\cite{WLL13}. Interestingly, the recent
GRB~130427A has been observed with unprecedented detail at high
energies, on timescales as large as $10^5\,$s. The spectrum of the
extended emission for this burst reveals a clear distinct component
above a few~GeV, which is well reproduced by a
synchrotron-self-Compton component~\cite{Tea13,LWW13}. The light curve
and the spectrum of this gamma-ray burst thus conform rather well to
the above expectations.

\subsection{Protons}
One can also compute the maximum energy of protons accelerated at the
external shock wave, by balancing the acceleration timescale with the
dynamical timescale. One then finds maximal energies of order
$10^{16}\,$eV, the exact prefactor depending on whether one assumes
that scattering is dominated by the small-scale field or by the large
scale field upstream of the shock~\cite{GA99,EP11,Bea12,PPL13}. Among
others, this confirms that the external shock wave of gamma-ray bursts
cannot produce particles in the ultra-high energy range.

The difficulty of pushing particles to extreme energies in
ultra-relativistic shock waves is directly associated to the
small-scale nature of the turbulence and the associated energy scaling
of the scattering rate, which implies $t_{\rm acc}\,\propto\, t_{\rm
  g}\,r_{\rm g}/\lambda_{\delta B}\,\gg\,t_{\rm g}$. Invoking a large
external magnetic field would not help, because the dynamic range
$\gamma_{\rm max}/\gamma_{\rm min}$ available to acceleration scales
at most as $\delta B/B\,\sim\,(\epsilon_B/\sigma)^{1/2}$, see the
discussion around Eq.~(\ref{eq:acc1}).

There are nevertheless loopholes in the above argument, which may lead
to potential acceleration sites of ultra-high energy cosmic
rays. First of all, the above discussion applies to ultra-relativistic
shock waves; for mildly relativistic shock waves, $\gamma_{\rm
  sh}\beta_{\rm sh}\,\sim\,{\cal O}(1)$, the shock may be sub-luminal
in a substantial part of parameter space, hence Eq.~(\ref{eq:acc1})
does not necessarily holds. This suggests that acceleration might take
place in large-scale, possibly self-amplified, turbulence, with a fast
scattering closer to the Bohm limit at high energies. Acceleration to
ultra-high energies at such shock waves has been proposed in gamma-ray
bursts~\cite{W95,GP04}, blazars~\cite{DRFA09} and in
trans-relativistic supernovae~\cite{Wea07,Bea08,Cea11,LW12}, although
the latter could push only heavy nuclei up to $10^{20}\,$eV. Another
loop-hole is the implicit assumption of a steady planar shock front;
if an additional source of turbulence or dissipation exists in the
vicinity of the shock, then one might circumvent the limitation of
ultra-relativistic acceleration. The termination shock of the wind of
the Crab pulsar, and other pulsar wind nebulae, provides a nice
example: with a shock magnetization $\sigma \,\gtrsim\, 10^{-2}$ and
shock Lorentz factor $\gamma_{\rm sh}\,\sim\,10^3-10^6$, the Crab
unexpectedly accelerates electrons at a Bohm-like rate, up to PeV
energies; yet, how acceleration takes place in such objects remains a
subject of debate, see Ref.~\cite{KLP09} for a detailed
discussion. One should thus question whether the termination shocks of
young pulsar winds, if they output ions, could not accelerate such
particles to ultra-high energies; work is in progress and preliminary
results are encouraging.

\section{Synchrotron spectra in micro-turbulence}
The small-scale nature of the turbulence is also likely to entail
departures from the standard synchrotron spectra of relativistic blast
waves. The possibility of a diffusive synchrotron regime, in which the
electron wanders over several coherence cells of the magnetic field
while the emission cone of aperture $\sim\,1/\gamma_e$ sweeps the line
of sight, has received much attaction,
e.g.~\cite{M00,FU10,KR10,MW11,Mea11,KAK13}. However it is now
recognized that such effects must remain very limited at relativistic
shock waves because the wiggler parameter $a\,\equiv\, e\delta
B\lambda_{\delta B}/(m_e c^2)\,\sim\,\gamma_{\rm
  min}\epsilon_{B,-2}^{1/2}$ with $\gamma_{\rm min}\,\sim\,\gamma_{\rm
  sh}\,m_p/m_e\,\gg\,1$ the typical Lorentz factor of the accelerated
electrons; $a\,\gg\,1$ means that the spectral power emitted follows
closely the standard synchrotron shape.

\begin{figure}[pb]
\centerline{\psfig{file=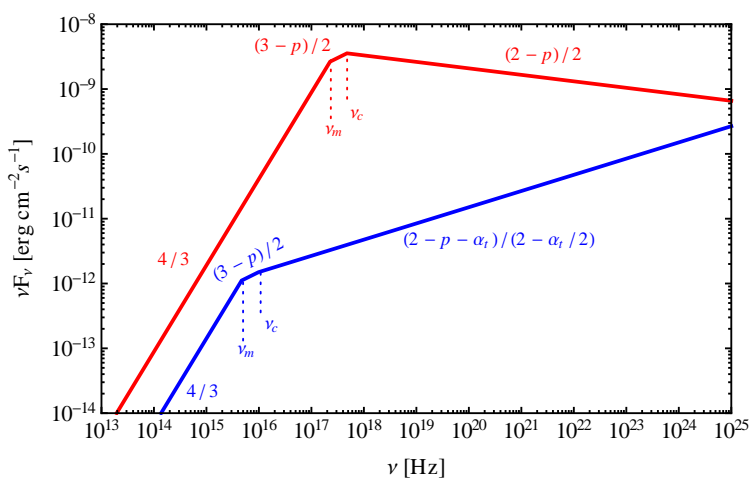,width=10cm}}
\vspace*{8pt}
\caption{Comparison of the synchrotron spectrum in a homogeneous
  turbulence of strength $\epsilon_B=10^{-2}$ (upper red curve) with a
  decaying microturbulence such that $\alpha_t=-0.8$ (lower blue
  curve), assuming in both cases dominant inverse Compton energy
  losses with $Y=3$. This case assumes $\gamma_{\rm b}=245$,
  $n=0.001\,$cm$^{-3}$, an injection distribution index $p=2.2$, and a
  cooling time of the electrons longer than the dynamical
  timescale. \label{fig:synch}}
\end{figure}

Nevertheless, the turbulence is expected to decay on short timescales
through phase mixing~\cite{GW99} and PIC simulations do confirm this
decay with a rough behavior $\epsilon_B\,\propto\,t^{\alpha_t}$,
$\alpha_t\,\sim\,-0.5$~\cite{Cea08,Kea09}, see also the discussion in
Ref.~\cite{L13}; in this equation, $t$ designs the time of injection
of the plasma element through the shock, in the comoving downstream
frame; thus, $t\,\sim\,3\Delta x/c$ where $\Delta x$ is the distance
to the shock front. This decay implies that along its cooling history,
an electron will cool in a magnetic field of decreasing strength,
which directly modifies the synchrotron spectral power of a single
electron. Furthermore, electrons of different Lorentz factors cool in
regions of different magnetic field strengths due to the energy
dependence of the cooling time. This has obvious consequences for the
synchrotron spectrum of a relativistic blast wave.  The departures
from the standard one-zone synchrotron spectrum have been addressed in
simplified limits in Refs.~\cite{RR03,D07} and calculated in detail in
Ref.~\cite{L13}, which has also quantified the impact on the light
curve of a decelerating blast wave.  Figure~\ref{fig:synch}
illustrates this departure by comparing the synchrotron spectrum of a
gamma-ray burst in the homogeneous model with $\epsilon_B=0.01$ (upper
red curve) and in a model with power-law decaying microturbulence,
$\epsilon_B\,=\,0.01\left(t/100\omega_{\rm p}^{-1}\right)^{-0.8}$
(lower blue curve) assuming in both cases that inverse Compton losses
dominate the cooling history with $Y=3$ for the Compton parameter
close to the shock front.

In this frame, it is interesting to note that the synchrotron modeling
of the extended high energy emission seen by the {\it Fermi-LAT}
instrument in several gamma-ray bursts systematically point to quite
low values
$\epsilon\,\sim\,10^{-6}-10^{-4}$~\cite{KBD09,KBD10,BDK11,Hea11,LW12}. As
discussed in Ref.~\cite{L13}, such values may actually attest of the
decay of the turbulence downstream of the shock; this would allow to
reconcile the afterglow models of gamma-ray bursts with the
predictions of PIC simulations, a long-standing problem in this
field. As an order of magnitude estimate, note that the blast width is
typically $10^7$ skin depths $c/\omega_{\rm p}$, so that the power-law
decay of $\epsilon_B$ down to $\sim 10^{-5}$, starting at $0.01$ some
$100c/\omega_{\rm p}$ away from the shock, indeed takes place on the
scale of the blast if $\alpha_t\,\sim\,-0.5$.

More quantitatively, one can build an approximate two-zone model of
the blast: a first zone close to the shock front with
$\epsilon_{B,+}\,\sim\,0.01$, in which the highest energy electrons
cool and produce the $>100\,$MeV emission; a second zone with
$\epsilon_{B,-}\,\ll\,\epsilon_{B,+}$, close to the contact
discontinuity, where electrons, which do not cool on a dynamical
timescale, output most of their synchrotron emission. This latter
region produces the low energy part of the synchrotron spectrum,
i.e. the radio, optical and at early times the X-ray region. One can
use the standard homogeneous synchrotron model for each zone and
adjust the model to the data to derive $\epsilon_{B,-}$. This
reconstruction is discussed in detail in Ref.~\cite{LLW13}. A
non-trivial aspect is the fact that the low-energy electrons cool
mostly through inverse Compton interactions, due to the low value of
$\epsilon_{B,-}$ and, in the case of X-ray emitting electrons,
Klein-Nishina suppression of the inverse Compton interactions is
significant. The cooling history of these electrons is thus not
trivial and this effect needs to be taken into account, otherwise
incorrect values for the microphysical parameters would be derived
from the observations. Furthermore, the afterglow model depends on 4
main parameters: 2 macrophysical (blast energy $E$ and external
density $n$) and 2 microphysical ($\epsilon_{B,-}$ and
$\epsilon_e\,\sim\,0.1$); the parameters $p$ and $k$, which
characterize respectively the injection spectral index and the
circumburst density profile, can be derived from the slopes of the
light curve. In order to derive the 4 parameters, one thus needs to
adjust the model to 4 different wavebands. For this reason,
Ref.~\cite{LLW13} has studied gamma-ray bursts afterglows that were
caught in the radio, optical, X-ray and at high energy
$>100\,$MeV. Once such a proper reconstruction of $\epsilon_{B,-}$ is
carried out, one finds quite remarkably, a coherent value of the decay
exponent $\alpha_t$ among the four bursts studied,
$-0.5\,\lesssim\,\alpha_t\,\lesssim\,-0.4$. Whether this result
pertains over a larger number of gamma-ray bursts remains to be seen,
of course. It is interesting to note that the recent GRB~130427 fits
well in this model, because its modelling leads to
$\alpha_t\,\simeq\,-0.45$. At the very least, these results raise the
exciting prospect of probing the physics of shock-generated Weibel
turbulence through the multi-wavelength afterglows of gamma-ray
bursts.

\section{Summary}
This papers summarizes recent progresses in our theoretical
understanding of particle acceleration at relativistic shock
waves. Emphasis is placed here on the intimate relationship that
appears to tie the generation of micro-turbulence in the precursor of
ultra-relativistic shock waves ($\gamma_{\rm sh}\,\gg\,1$) and the
efficiency of particle acceleration. The arguments reported here have
been developed at an analytical level and validated by
particle-in-cell simulations. This includes in particular the
development of acceleration over a broad dynamic range whenever the
magnetization of the upstream plasma~\cite{LP10}
$\sigma\,\lesssim\,\epsilon_B^2\,\sim 10^{-5}$, assuming
micro-turbulence has been excited by Weibel/filamentation or through a
current instability discussed by G.~Pelletier in this volume.

This paper then discusses two salient consequences of this
micro-turbulence: (1) the maximal energy of particle accelerated at
the shock, which departs from the Bohm scaling because the scattering
frequency in the micro-turbulence decreases with increasing energy;
(2) the impact of the evanescent nature of this micro-turbulence on
the afterglow spectra and light curves of gamma-ray bursts, where it
is argued in particular that the decay of this turbulence may well
have been indirectly observed in gamma-ray bursts with extended high
energy emission.



\end{document}